\documentclass{aipproc}
\layoutstyle{6x9}

\begin{document}
\title{Inflation after WMAP3}
\author{William H. Kinney}{
address={Dept. of Physics, The University at Buffalo, SUNY, Buffalo, NY}
}

\begin{abstract}
I discuss the current status of inflationary cosmology in light of the recent WMAP 3-year data release. 
The basic predictions of inflation are all supported by the data. Inflation also makes predictions which have not been well tested by current data but can be by future experiments, most notably a deviation from a scale-invariant power spectrum and the production of primordial gravitational waves. A scale-invariant spectrum is disfavored by current data, but not conclusively. Tensor modes are currently poorly constrained, and slow-roll inflation does not make an unambiguous prediction of the expected amplitude of primordial gravitational waves. A tensor/scalar ratio of $r \simeq 0.01$ is within reach of near-future measurements.
\end{abstract}

\date{\today}

\keywords{inflation,CMB}
\classification{98.80.Cq}

\maketitle

\section{Introduction: The Inflationary Model Space}

Inflation \cite{Guth:1980zm} has emerged as the most successful model for understanding the physics of the very early universe. Inflation in its most general form consists of a period of accelerating expansion, during which the universe is driven toward flatness and homogeneity. In addition, inflation provides a mechanism for generating the initial perturbations which led to structure formation in the universe. The key ingredient of this cosmological acceleration is negative pressure, or a fluid with a vacuum-like equation of state $p \sim - \rho$. In order for inflation to end and the universe to transition to the radiation-dominated expansion necessary for primordial nucleosynthesis, this vacuum-like energy must be dynamic, and therefore described by one or more order parameters with quantum numbers corresponding to vacuum, {\em i.e.} scalar fields. In the absence of a compelling model for inflation, it is useful to consider the simplest models, those described by a single scalar order parameter $\phi$, with potential $V(\phi)$ and energy density and pressure for a homogeneous mode of
\begin{equation}
\rho = {1 \over 2} \dot\phi^2 + V\left(\phi\right),\ 
p = {1 \over 2} \dot\phi^2 - V\left(\phi\right).
\end{equation}
The negative pressure required for inflationary expansion is achieved if the field is slowly rolling, $\dot\phi^2 \ll V(\phi)$, so that the potential dominates. During inflation, quantum fluctuations on small scales
are quickly redshifted to scales much larger than the horizon size, where they
are ``frozen'' as perturbations in the background
metric. The metric perturbations created during inflation are of two types, both of
which contribute to anisotropy in the Cosmic Microwave Background (CMB):
scalar, or {\it curvature} perturbations, which couple to the stress-energy of
matter in the universe and form the ``seeds'' for structure formation, and
tensor, or gravitational wave perturbations, which do not couple to matter. Different inflationary models can then be constructed by specifying different choices for the potential $V(\phi)$. In turn, different choices for $V(\phi)$ predict different spectra for primordial fluctuations in the universe, and precision observations can shed light on the physics relevant during the inflationary epoch. The mapping of the inflationary parameter space onto the observable parameter space is widely covered in the literature, and the reader is referred to Refs \cite{Kinney:2003xf,Lyth:2007qh} for reviews. The basic predictions of slow-roll inflation models are straightforward to summarize: generically, Gaussian, adiabatic scalar and tensor fluctuations will be created during inflation with approximately power-law spectra. The scalar spectrum is conventionally parameterized in terms of a spectral index $n$ as
\begin{equation}
P_{R} = k^{n - 1},
\end{equation}
and the tensor spectrum as 
\begin{equation}
P_{T} = k^{n_T}.
\end{equation}
Absolute normalization of the power spectra is governed by an adjustable (and usually fine-tuned) parameter in the inflationary potential, but the {\it ratio} of the normalizations is not,
\begin{equation}
\label{eq:consistency}
r \equiv {P_{T} \over P_{R}} = -8 n_{T},
\end{equation}
known as the {\it consistency condition}, which applies to single-field models of inflation. The tensor/scalar ratio $r$ also tells us the equation of state during inflation, $r = 16 \epsilon$,
where the {\it slow roll parameter} $\epsilon$ is related to the equation of state by
\begin{equation}
p = \rho \left[(2/3) \epsilon - 1\right],
\end{equation}
and $\epsilon$ is in turn related to the inflationary potential by
\begin{equation}
\epsilon \simeq {M_{\rm Pl}^2 \over 2} \left({V'\left(\phi\right) \over V\left(\phi\right)}\right)^2,
\end{equation}
where $M_{\rm Pl} = 1 / \sqrt{8 \pi G}$ is the reduced Planck mass. Because of the consistency condition (\ref{eq:consistency}), the tensor spectral index $n_{T}$ is not an independent parameter. The parameters relevant for distinguishing among simple slow-roll inflation models are $n$ and $r$, which can be determined from current observational data. \footnote{More complex inflation models could potential produce additional signals such as non-Gaussianity or running ({\it i.e.} scale dependence) of the spectral index $n$. Current data provide no statistically significant evidence of either \cite{Spergel:2006hy,Kinney:2006qm}, and I do not consider the possibility further here.} In the next section, I will discuss constraints on the inflationary parameter space from the most precise existing measurement of the CMB anisotropy, the WMAP 3-year data set \cite{Spergel:2006hy}. The results I report here are covered in detail in WHK, Kolb, Melchiorri, and Riotto, Ref. \cite{Kinney:2006qm} (KKMR).

\section{Results from the WMAP 3-year data set}
\label{sec:WMAP3}

\begin{figure}
\resizebox{.8\textwidth}{!}{\includegraphics{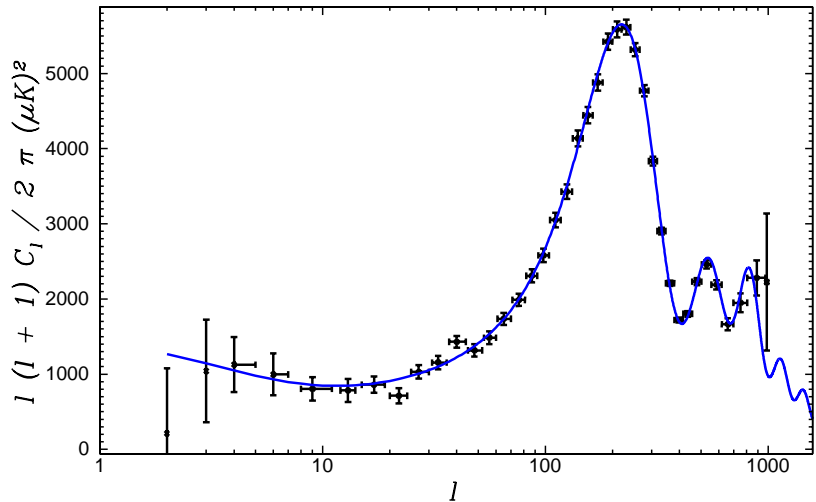}}
\caption{\label{fig:spectrum} An adiabatic power-law spectrum with $n = 0.99$ and $r = 0.26$, compared with the WMAP3 data.}
\end{figure}

The WMAP 3-year data set (WMAP3) \cite{Spergel:2006hy} represents the most sensitive all-sky map of the CMB made to date, and places strong constraints on the age, contents, and geometry of the universe. The result which will be of primary interest here is that WMAP3 is remarkably consistent with the predictions of inflation. Inflation generically predicts a universe with a geometry exponentially close to flat, a result which is consistent with the curvature constraint from WMAP3. Inflation in its simplest slow-roll realizations also makes a very specific prediction about the primordial spectrum of density perturbations. Slow roll inflation models predict a primordial power spectrum which is: (a) Gaussian, (b) adiabatic, and (c) close to (but not exactly) a scale-invariant power law. Different models predict different spectral indices or degrees of running, but these three basic properties are robust predictions of slow-roll inflation.  Figure \ref{fig:spectrum} shows the prediction of a best-fit Gaussian, adiabatic power law along with the WMAP3 data points for the $C_{\ell}$ spectrum for the temperature anisotropy autocorrelation. The agreement between the inflationary prediction and the data is remarkable: WMAP3 resolves the first three acoustic peaks characteristic of adiabatic fluctuations, and the overall spectrum is well-fit by a scale-invariant power law. 

Should this correspondence between theory and data be considered in some sense a confirmation of the inflationary paradigm? It is worth emphasizing that the data did not have to turn out this way: a non-flat universe or primordial perturbations from cosmic strings would have left a radically different signature in the CMB anisotropy spectrum. This agreement extends to more than the temperature anisotropy: adiabatic fluctuations result in specific correlations between the temperature anisotropy and the polarization of the CMB which are also an excellent fit to the data.\footnote{A good discussion of this issue can be found in Ref. \cite{Peiris:2003ff}.} There is strong empirical evidence to support inflation as {\it a} theory of the very early universe, but it is perhaps premature to conclude that existing evidence points to inflation as {\it the} theory of the very early universe.  A scale-invariant, adiabatic perturbation spectrum was proposed many years before the development of inflation by Harrison and Zel'dovich based on symmetry principles alone, but what they did not propose was a mechanism for generating correlations in the perturbations on superhorizon scales. Superhorizon correlations are a key signature of inflationary physics, and have been argued to be definitive evidence for inflation \cite{Spergel:1997vq}. Nonetheless, alternatives to inflation have been proposed, the best known of which is the Ekpyrotic/Cyclic scenario \cite{Khoury:2001zk}. The true viability of this model as an alternative to inflation is controversial \cite{Linde:2002ws,Martin:2002ar}, but the message remains that what appear to be acausal correlations in the CMB can be produced either by inflationary expansion, or by the introduction of extra dimensions, as is common in braneworld scenarios. Regions which appear to be causally disconnected on a $3+1$ brane may not be so in the higher-dimensional bulk. Acausal correlations may also be induced by a variation in the speed of light \cite{Albrecht:1998ir}, or possibly by a Hagedorn phase in the early universe \cite{Brandenberger:2007qi}. There is considerable debate as to whether inflation in its most general sense is even falsifiable \cite{Barrow:1997fp}, but particular slow roll models can certainly be ruled out by existing and future data \cite{Dodelson:1997hr,Kinney:1998md}. 

One observational result which would increase confidence that inflation is the correct model for the early universe would be the {\it exclusion} of the simple Harrison-Zel'dovich (HZ) model to a high degree of confidence. Here I will define the HZ model to be a scale-invariant ($n = 1$) spectrum consisting purely of adiabatic density fluctuations, with no tensor component present. Therefore, the two ways in which HZ might be excluded in the data are:
\begin{itemize}
\item{A measurable deviation from scale invariance ($n \neq 1$)}
\item{A detectable contribution to the CMB anisotropy from a background of primordial gravitational waves ($r \neq 0$).}
\end{itemize}
I consider each separately below. 

\subsection{Deviation from scale invariance}

\begin{figure}
\resizebox{.8\textwidth}{!}{\includegraphics{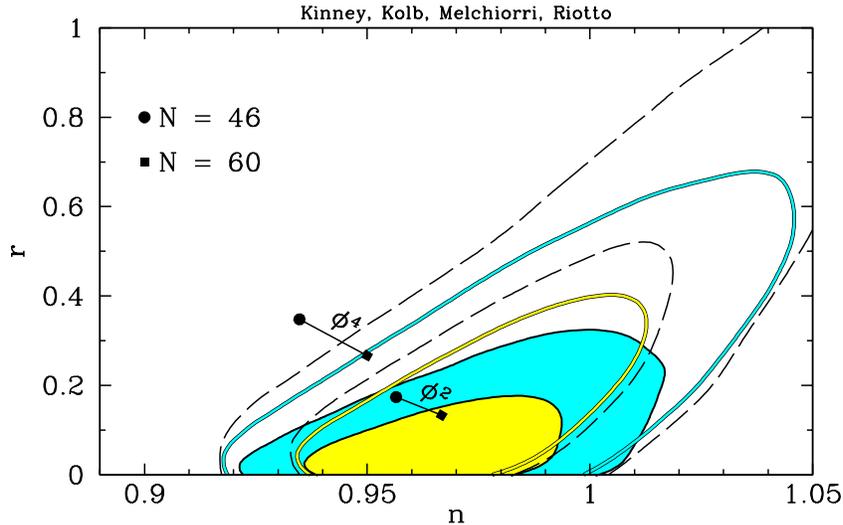}}
\caption{\label{fig:zoonorun} Allowed regions in the $(n,r)$ parameter space. The dashed curves show the 68\% and 95\% confidence regions from the analysis released publicly by the WMAP team, which does not include a HST prior on $h$. The open shaded curves are the 68\% and 95\% confidence regions from the KKMR analysis for the WMAP3 data set. The inner filled contours are the the 68\% and 95\% confidence regions for WMAP3 taken in combination with data from the Sloan Digital Sky Survey. The lines labeled $\phi^2$ and $\phi^4$ are the predictions of inflationary models with the corresponding potentials. The HZ model is at $n = 1$ and $r = 0$.}
\end{figure}

In order to realistically determine which regions of the inflationary parameter space $r$ and $n$ are consistent with the data, it is necessary to take into account possible degeneracies with other cosmological parameters, such as the baryon density $\Omega_{\rm b}$ or the reionization optical depth $\tau$. The technique which has become standard is a Bayesian parameter analysis using a Monte Carlo Markov Chain for numerical efficiency \cite{Lewis:2002ah}. The results I report here are from KKMR \cite{Kinney:2006qm}, where we performed an analysis varying the following seven parameters:
\begin{itemize}
\item{Baryon density $\Omega_{\rm b} h^2$}
\item{Cold Dark Matter density $\Omega_{\rm c} h^2$}
\item{Angular diameter distance at decoupling $\theta$}
\item{Reionization optical depth $\tau$}
\item{Power spectrum normalization $A_{\rm s}$}
\item{Scalar spectral index $n$}
\item{Tensor/scalar ratio $r$}
\end{itemize}
The overall curvature is fixed to zero by adjusting the Dark Energy density such that $\Omega_{\rm total} = 1$, and the inflationary consistency condition (\ref{eq:consistency}) is assumed. The equation of state of the Dark Energy is fixed at $w = -1$. A tophat age prior of $t_0 = 10 - 20\ {\rm Gyr}$ is assumed, as well as a HST prior on the Hubble Constant of $h = 0.72 \pm 0.08$. Figure \ref{fig:zoonorun} shows the allowed regions in the $(n,r)$ plane for WMAP3 alone and WMAP3 in combination with the Sloan Digital Sky Survey (SDSS) \cite{Tegmark:2003uf} data set. WMAP3 places strong constraints on the inflationary parameter space. What is probably the simplest possible inflation model, $V(\phi) = m^2 \phi^2$, is fully consistent with existing data. Not so a model with $V(\phi) = \lambda \phi^4$. Such a potential is marginally consistent with the WMAP3 data when taken alone, but is ruled out to well better than 95\% confidence by WMAP3 in combination with SDSS. Also ruled out at the 95\% level are tree-level hybrid models of the type originally suggested by Linde \cite{Linde:1993cn}, which predict a blue spectrum $n > 1$ and negligible tensor component, $r \simeq 0$ \cite{Kinney:2006qm}. 

The Harrison-Zel'dovich spectrum, however, is inside the 95\% confidence contour, a result which is robust with respect to choice of a prior on the Hubble constant, and with respect to choice of data set (inclusion of SDSS). This conclusion is at odds with statements made in the literature that the HZ spectrum is ruled out to better than 99\% confidence, for example by Kamionkowski in Ref. \cite{Kamionkowski:2007wv}. The difference in quoted statistics depends on whether or not one includes $r$ in the parameter set allowed to vary in the Bayesian fit: a six-parameter fit with a prior of $r = 0$ produces much tighter error bars on $n$. This is an example of the importance of considering priors when drawing conclusions from a Bayesian analysis, a subject discussed lucidly by Parkinson, Mukherjee, and Liddle in Ref. \cite{Parkinson:2006ku}. 
The bottom line is that the HZ model is disfavored by the WMAP3 constraint on the scalar spectral index, but it is very difficult to argue that the evidence is conclusive. Future measurements such as the Planck satellite \cite{:2006uk} will make possible significantly improved constraints on $n$, and will be capable of definitively distinguishing the HZ spectrum from a spectrum with $n \leq 0.98$. 

\subsection{Gravitational wave background}

Inflation predicts not just the generation of curvature (scalar) perturbations in the early universe, but also the generation of gravitational wave (tensor) perturbations. If the tensor component is large enough, it will be detectable by upcoming CMB measurements. Current limits on the tensor contribution to the CMB spectrum are extremely weak, with an upper limit of $r < 0.6$ at 95\% confidence  for a seven-parameter fit with no running of the scalar spectral index, and $r < 1.1$ for an eight parameter fit with running included \cite{Kinney:2006qm}. A substantial increase in sensitivity will be required of future measurements to place a meaningful limit on $r$. In the near term, future CMB measurements could realistically probe the tensor/scalar ratio down to $r \simeq 0.01$ \cite{Kinney:1998md}, while in the more distant future, direct detection of primordial gravitational waves may be feasible  to a level of $r \sim 0.0001$ \cite{Smith:2005mm}. With such ambitious observational efforts either in progress or on the drawing board, there is considerable interest in the question of what inflation predicts for the amplitude of primordial gravitational waves. There have been several approaches to addressing this question proposed in the literature. Lyth showed that the tensor/scalar ratio $r$ can be related to the variation in the inflaton field $\Delta \phi$ by the inequality \cite{Lyth:1996im}
\begin{equation}
\Delta\phi > 0.46 M_{\rm Pl} \sqrt{r \over 0.07}.
\end{equation}
This is significant because effective field theory arguments ({\it e.g.} from stringy model building) suggest that the field variation should be small compared to the Planck scale, $\Delta\phi \ll M_{\rm Pl}$, resulting in a strongly suppressed tensor amplitude. However, models such as Natural Inflation \cite{Freese:1990rb,Savage:2006tr} and N-flation \cite{Dimopoulos:2005ac} can achieve $\Delta\phi \sim M_{\rm Pl}$ in a technically natural way, so this constraint does not appear to be inescapable. 

A more recent argument is that of Boyle, Steinhardt, and Turok who use a counting argument to conclude that a suppressed tensor/scalar ratio requires a highly fine-tuned potential \cite{Boyle:2005ug}. They conclude that, in the absence of fine tuning, a red spectrum (as favored by WMAP3) results in an observably large tensor/scalar ratio, $r > 0.01$. This argument is also severely weakened by the existence of a counterexample, that of an ``inverted'' potential with a suppressed mass term, which can be approximated for small $\phi$ by the leading-order behavior
\begin{equation}
\label{eq:phi4}
V(\phi) = V_0 - \lambda \phi^4.
\end{equation}
The key property of such potentials is that the Planck scale $M_{\rm Pl}$ cancels in the expressions for the power spectrum normalization and spectral index \cite{Knox:1992iy,Kinney:1995cc},
\begin{eqnarray}
&&P_{R} \sim \lambda\cr
&&n = 1 - {3 \over N},
\end{eqnarray}
where $N = [46,60]$ is the number of e-folds of inflation. Therefore, inflation can take place at an arbitrarily low energy scale $V_0$ and still satisfy observational constraints. But since the tensor amplitude is $P_{T} \propto V_0 / M_{\rm Pl}^4$, a low energy scale means suppressed tensors. A potential of the form (\ref{eq:phi4}) would be labeled as unacceptably fine-tuned by the counting procedure of Ref. \cite{Boyle:2005ug}, but would certainly not be considered fine-tuned by any definition familiar to particle physicists: a potential of the form (\ref{eq:phi4}) is characteristic of scalar field potentials generated by radiative corrections, for example the Coleman-Weinberg model \cite{Coleman:1973jx}. An example of a fully-formed inflation model which meets the criteria for a successful model outlined in Ref. \cite{Boyle:2005ug}, {\it i.e.} a potential which is bounded below, stable with respect to radiative corrections, and coupled to fermions for successful reheating, can be found in Ref. \cite{Kinney:1995cc}. 

In summary, there are theoretical arguments as to why one might expect either outcome for $r$: field theory based tuning arguments favor unobservably small $r$, and slow-roll based tuning arguments favor $r > 0.01$, in the range accessible to observation. All of these arguments contain large loopholes, leaving the issue of the tensor amplitude from inflation (and in the real universe, whether inflationary or not) an open, intrinsically {\it observational} question. The only way to find out the answer is to go out and look.\footnote{The author acknowledges a bet with Latham Boyle of one bottle of scotch, brand to be determined by the winner, that the tensor/scalar ratio will turn out to be $r < 0.01$, and would be delighted to lose the bet.}

\section{Conclusions}

The milestone WMAP measurement is the first single, self-contained data set capable of placing meaningful constraint on the inflationary model space. Inflation has passed the test with flying colors. The basic predictions of the inflationary model are all supported by the data: a flat universe with Gaussian, adiabatic nearly scale-invariant perturbations. No other model explains these properties of the universe with such simplicity and economy. Inflation also makes predictions which are have not been well tested by current data but {\it can} be by future experiments, most notably a deviation from a scale-invariant spectrum and the production of primordial gravitational waves. The scale-invariant spectrum is disfavored by current data, but not conclusively. Tensor modes are currently poorly constrained, but a tensor/scalar ratio of $r \simeq 0.01$ is within reach of near-future measurements.\cite{Kinney:1998md} A detection of primordial gravitational radiation would provide strong evidence for a period of inflation in the very early universe. Unfortunately, inflation models do not make an unambiguous prediction of the expected amplitude of primordial gravitational waves. The issue will likely only be resolved by observation. 

\smallskip
This research is supported in part by the National  Science  Foundation under  grant NSF-PHY-0456777.

\end{document}